\documentclass[preprint,aps,nofootinbib,tightenlines, byrevtex]{revtex4}

\usepackage{amsfonts}
\usepackage[dvips,final]{graphicx}

\usepackage{multirow}                     
\usepackage{xspace}                       

\newcommand{\ii}{\mathrm{i}}

\newcommand{\mpi}{\ensuremath{m_\pi}}

\newcommand{\EFTNoPion}{EFT(${\pi\hskip-0.55em /}$)\xspace}

\newcommand{\LambdaNoPion}{\ensuremath{\Lambda_{\pi\hskip-0.4em /}}}

\newcommand{\calH}{\mathcal{H}}

\begin{document}
\title{Application of EFT at Thermal Energies}
\author{H. Sadeghi}\email{H-Sadeghi@araku.ac.ir}
 \affiliation{Department of Physics, University of Arak, P.O.Box 38156-879, Arak,
 Iran.}

 \vspace{4cm}
\begin{abstract}
We have been evaluated some observables of n-d systems  by using
pionless Effective Field Theory(\EFTNoPion) and insertion of the
three-body force up to next-to-next to leading order(N$^2$LO). The
evaluated data has been compared with experimental and the
three-nucleon calculation of the total cross section with modern
realistic two- and three-nucleon forces AV18/UrbIX potential models
calculations.
\end{abstract}

\maketitle
\section{Introduction}
Formation of light nuclei from nucleons, help us for understanding
of nuclear structure in Big-Bang Nucleosynthesis(BBN). For these
calculations EFT is one of the most important tools. Results with
this theory, in relevant range of energies, for
two-body~\cite{chen99} and recently for three body
systems~\cite{3stooges_boson} show very good correspondence in
comparison with the experimental and evaluated results. Recently, we
applied \EFTNoPion to find numerical results for some astrophysical
observables at thermal energies~\cite{Sadeghi1,Sadeghi2,Sadeghi3}.
For these calculations all particles but nucleons are integrated
out. Three-nucleon forces are added up to N$^2$LO for cut-off
independent results.

In this paper, we briefly review calculation of some BBN
observables.  The cross-section for neutron-deuteron capture
process~\cite{Sadeghi2} and the photon polarization
parameter($R_c$)~\cite{Sadeghi3} are determined. These converges
order by order in low energy expansion and also is cut-off
independent at this order. We have been also studied
$\gamma{^3H}\rightarrow nd$ process near threshold by using
\EFTNoPion~\cite{Sadeghi4}. The evaluated cross section has been
compared with experimental and the three-nucleon photodisintegration
calculation of the total cross section with the modern realistic
two- and three-nucleon forces AV18/UrbIX potential models
calculations.
\section{Formalism}
\begin{figure}[!htb]
\begin{center}
 \includegraphics*[width=.8\textwidth]{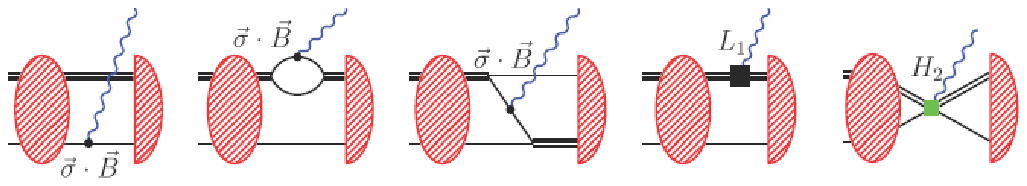}
 \caption{ The Faddeev equation for $Nd$-scattering and adding photon-interaction to the Faddeev equation
 up to  N$^2$LO. Wavy line shows photon and small
circles show magnetic photon interaction. See Ref.[3,4] for
notations.}
  \label{fig1}
\end{center}
\end{figure}
The $^2S_{1/2}$ describes the preferred mode for
$nd\rightarrow{^3H}\gamma$ and $pd\rightarrow{^3H}e\gamma$
processes. The integral equation describing neutron-deuteron
scattering has been discussed before~\cite{3stooges_boson}. The
integral equation is solved numerically by imposing a cut-off
$\Lambda$. In that case, a unique solution exists in the
$^2S_{1/2}$-channel for each $\Lambda$ and $\calH=0$, but no unique
limit as $\Lambda\to\infty$. Bedaque et al.~\cite{3stooges_boson}
showed that the system must be stabilized by a three-body force.
which absorbs all dependence on the cut-off as $\Lambda\to\infty$.

The neutron-deuteron $J=1/2$ phase shifts $\delta$ is determined by
the on-shell amplitude $t_t(k,k)$, multiplied with the wave function
renormalisation
\begin{equation}
T(k)=Z t_t(k,k)=\frac{3\pi}{M}\frac{1}{k \cot\delta-\ii k}\;\;.
\end{equation}
 For both possible magnetic dipole transitions with
 $ J^{P}=\displaystyle\frac{1}{2}^+$ (amplitude $g_1$) and
 $J^{P}=\displaystyle\frac{3}{2}^+$ (amplitude $g_3$) we can write:
$$g_1:~~t^\dagger(i\vec D\cdot\vec{e^*}\times\vec k+\vec\sigma\times\vec
D\cdot\vec{e^*}\times\vec k)N ~~and~~ g_3:~~t^\dagger(i\vec
D\cdot\vec{e^*}\times\vec k+\vec \sigma\times\vec
D\cdot\vec{e^*}\times\vec k)N\;. \label{eq:as8}$$ and the
$nd\rightarrow ^3H\gamma$ cross section at very low energy is given
by
\begin{equation}\label{crosssection}
  \sigma=\frac{2}{9}\frac{\alpha}{v_{rel}}\frac{p^3}{4M^2_N}\sum_{iLSJ}
  [{|\widetilde{\chi}^{LSJ}_i|}^2]\;,~~where~~
  \widetilde{\chi}^{LSJ}_i=\frac{\sqrt{6\pi}}{p\mu_N} \sqrt{4\pi}
  {\chi^{LSJ}_i}\;,
\end{equation}
with $\chi$ stands for either E or M and $\mu_N$ is in nuclear
magneton and p is momentum of the incident neutron in the center of
mass.
\section{Results and Discussion}
\begin{table}[!htb]
\caption{Comparison between different theoretical and experimental
results for $nd\rightarrow{^3H}\gamma$ and $R_c$ . Last rows shows
our EFT result up to N$^2$LO.} \label{tab1}
\begin{center}
\begin{tabular}{c||c|c}
  Theory  & $nd\rightarrow{^3H}\gamma$(mb) & $R_c$ \\
\hline
      AV18/IX(IA+MI+MD+$\Delta$)~\cite{Viviani1}  & 0.631 & -0.469   \\
      AV18/IX (gauge inv. + 3N-current) ~\cite{Marcucci} & 0.578 & -0.476    \\
   EFT(N$^2$LO)~\cite{Sadeghi2} & 0.503   &  -0.412   \\
   Experiment~\cite{Jurney}  & $0.503\pm 0.003$ & $-0.42\pm0.03$  \\
\hline
\end{tabular}
\end{center}
\end{table}
We applied \EFTNoPion to find numerical results for some
astrophysical observables in low energies. At very low energies, the
interactions between nucleons can be described only by point-like
interactions. All particles but nucleons are integrated out and one
can identify a small, dimensionless parameter
$Q=\frac{p_\mathrm{typ}}{\LambdaNoPion}\ll 1$, where
$\LambdaNoPion\sim\mpi$ is the typical momentum scale at which the
one pion exchange is resolved and \EFTNoPion must break down.
Incident thermal neutron energies  have been considered for this
capture process. At these energy our calculation is dominated by
only $S$-wave state and magnetic transition $M_1$ contribution. The
$M_1$ amplitude is calculated up to N$^2$LO. Relevant diagrams has
been shown in Fig.~\ref{fig1}. The triton binding energy and nd
scattering length in the triton channel have been used to fix them.
Numerical data for these calculations can be find in the
Refs.~\cite{Sadeghi1,Sadeghi2,Sadeghi3}.

The cross-section for nd capture process and $R_c$ are in total
determined as $0.503\pm 0.003$ mb  and $-0.412\pm0.003$,
respectively. In table {1}, the evaluated data has been compared
with experimental and the total cross section with modern realistic
three-nucleon forces AV18/UrbIX potential models calculations.  The
cross-section for triton photodisintegration process including of
$M_1$ and $E_2$ contributions  at 12 MeV is 0.866 mb in comparison
with 0.882 mb result calculated by using AV18 and UrbIX with
three-nucleon forces~\cite{Skibinski}. At the higher energies the
spreads are negligible which points to a certain stability of the
results and helps to identify 3N force effects. Our results
converges order by order in low energy expansion, is cut-off
independent at this order and has a systematic error which is now
smaller than the experimental error bar.

\section{acknowledge} My thanks to the EFB(20) organizers for his
support and the warm welcome during my stay in Pisa. This work has
been Supported by University of Arak.

\end{document}